\def\be{\begin{equation}}
\def\ee{\end{equation}}
\def\ba{\begin{array}}
\def\ea{\end{array}}

\newcommand{\bra}[1]{\langle #1|}
\newcommand{\ket}[1]{|#1\rangle}
\newcommand{\tr}{\mbox{Tr}}

\documentclass[aps,amsmath,amssymb,amsfonts,showpacs,twocolumn,pra]{revtex4}
\usepackage{graphicx}
\def\qed{\leavevmode\unskip\penalty9999 \hbox{}\nobreak\hfill
     \quad\hbox{\leavevmode  \hbox to.77778em{%
               \hfil\vrule   \vbox to.675em%
               {\hrule width.6em\vfil\hrule}\vrule\hfil}}
     \par\vskip3pt}

\begin{document}

\title{Genuine Multipartite Entanglement of Superpositions}
\author{Zhihao Ma}
\affiliation{Department of Mathematics, Shanghai Jiaotong
University, Shanghai, 200240, China}
\affiliation{ Department of Physics and Astronomy, University College
London, Gower St., WC1E 6BT London, United Kingdom}

\author{Zhihua Chen}
\affiliation{Department of Science, Zhijiang college, Zhejiang
University of technology, Hangzhou, 310024, China}
\affiliation{Centre for Quantum Technologies, National University of
Singapore, 3 Science Drive 2, 117543 Singapore,}

\author{Shao-Ming Fei}
\affiliation{School of Mathematical Sciences, Capital Normal University,
Beijing 100048, P. R. China}
\affiliation{Max-Planck-Institute for Mathematics in the
Sciences, 04103 Leipzig, Germany}

\begin{abstract}

We investigate how the genuine multipartite entanglement is distributed among the components of superposed states.
Analytical lower and upper bounds for the usual multipartite negativity and the genuine multipartite entanglement negativity
are derived. These bounds are shown to be tight by detailed examples.

\end{abstract}


\maketitle

\section{Introduction}

As a corner stone of quantum mechanic \cite{Shankar}, the superposition principle plays key roles in the
applications to quantum information processing such as
quantum factorization algorithm \cite{Shor}, and is tightly related to some novel quantum phenomena such as
in Schr\"odinger cat paradox and quantum no-cloning theorem \cite{Nocloningtheorem}.
The existence of superposed quantum states has been experimentally
demonstrated by using photons \cite{Deleglise}, atoms \cite{Hammerer} and even viruses \cite{Cirac}.
Experiments have been also designed to study the wave-particle duality according to the superposition of wave
and particle \cite{Tang}, which shed new light in understanding the Bohr's principle of complementarity
and quantum mechanics as well.

On the other hand, as a novel phenomenon in composite quantum system, entanglement is a distinctive feature of
quantum mechanics and has intrinsic connections with many fundamental
problems in quantum mechanics \cite{Horodecki09,Guhne09,Fuchs11}.
A natural question raised is then what happens to the superposition of entanglement.

In \cite{Linden} Linden, Popescu and Smolin first studied the evolution law of entanglement of superposition.
They observed that the superposition of two separable states can give rise to an entangled one, while the superposition of two
entangled states can result in a separable one.
Since then the entanglement of superposition has been extensively studied for both bipartite and
multipartite systems \cite{Yu,Niset,Ou,Cavalcanti,Gour07,Song,Gour08,Ma11,Akhtarshenas11,Parashar,Maqic}.
However, so far there is no result about genuine multipartite entanglement (GME) of superpositions,
although genuine multipartite entangled states have been proved to be vital in carrying out many fundamental quantum
information processing tasks.

We will focus on the superposition of genuine multipartite entanglement in terms of
the GME measure which characterizes the global entanglement of a quantum system.
The GME is quite different from the usual multipartite entanglement.
A usual entangled state may be not genuine multipartite entangled. A genuine multipartite entangled state
is not separable under any bipartite partitions. There are different classes of multipartite entangled states.
For instance, for three-qubit states, there exist two classes of GME states, namely, GHZ state and W state
\cite{three-pure,three-mix}, which are not equivalent under local unitary transformations.
Compared with usual the entanglement, GME displays more complicated structures and bears
some special advantages. They are the key resources of measurement-based quantum computing \cite{Briegel}
and high-precision metrology \cite{Giovannetti}. They also play significant roles in quantum phase
transitions \cite{de Oliveira,Bruss}.

For three-qubit systems, a crucial measure for GME is the so-called three-tangle \cite{three},
which is a polynomial invariant that quantifies the genuine tripartite entanglement contained in a pure three-qubit state.
Three-tangle is introduced from the monogamy relation of tripartite entanglement. It is the first milestone
towards a systematic treatment of GME. It was found that for rank-2 mixed states, e.g., GHZ-state mixed with the W-state,
the three-tangle of superposed state is completely determined by the three-tangle of  superposition of the
GHZ-state and the W-state \cite{three-tan1,three-tan2}.

In the present work, we give a systematic investigation on the GME of arbitrary superposed states
by using a generalization of the concurrence \cite{Mintert05PR,Wootters98,af} which has close connection with the entanglement measure
negativity. Based on the generalized concurrence, we define two tripartite entanglement measures,
one is for usual tripartite entanglement, i.e., it quantify all the entanglement in a three-qudit state,
another is a GME measure quantifying the genuine tripartite entanglement. We then apply the two measures
to study entanglement in superpositions of two tripartite pure states of arbitrary dimension.
Interestingly we find that, for the superpositions of GHZ state and W state, our upper bound
always gives the exact value of its GME measure.

We first recall two widely used entanglement measures for bipartite quantum states.
Let $\mathcal{H}_{A}$ and $\mathcal{H}_{B}$ be Hilbert spaces of dimension
$m$ and $n$, respectively. The \emph{concurrence} of a pure bipartite state $\rho_{AB}=|\psi\rangle
\langle\psi|$ in $\mathcal{H}_{A}\otimes\mathcal{H}_{B}$ is defined as
$C(|\psi\rangle):=\sqrt{2\left(  1-Tr\rho_{A}^{2}\right)}$ \cite{Mintert05PR}. We denote by
$\rho_{\gamma}$, $\gamma=A,B$, the reduced density operators. It is well-known that
a pure state is separable if and only if its concurrence is zero.

Let $L_{\alpha}= (\ket{i}\bra{j}-\ket{j}\bra{i})/\sqrt{2}$ denote the $m(m-1)/2$ generators of $SO(m)$ on $\mathcal{H}_{A}$,
and $S_{\beta}$ the $n(n-1)/2$ generators of $SO(n)$ on $\mathcal{H}_{B}$.
Then the square of the concurrence can be rewritten as
\begin{equation}
C^{2}(|\psi\rangle)=\sum_{\alpha,\beta=1}
^{D_{1},D_{2}}|C_{\alpha\beta}|^{2}, \label{aa}%
\end{equation}
where $D_{1}=m(m-1)/2$, $D_{2}=n(n-1)/2$, $C_{\alpha\beta}%
=\langle\psi|\widetilde{\psi}_{\alpha\beta}\rangle$, $|\widetilde{\psi}%
_{\alpha\beta}\rangle=J^{1|2}_{\alpha\beta}|\psi^{\ast}\rangle$, with
$J^{1|2}_{\alpha\beta}=(L_{\alpha}\otimes S_{\beta})$ \cite{Akhtarshenas}.
(\ref{aa}) is a form of $\ell_{2}$-norm.

For a pure state $\rho=|\psi\rangle\langle\psi|$, if the
eigenvalues of $\rho_{A}$ are $\lambda_{1},...,\lambda_{m}$, $\lambda_{1}%
\geq...\geq\lambda_{m}$, then $C^{2}(|\psi\rangle)=\sum_{i,j=1}^{m}\lambda_{i}\lambda_{j}$.
An $\ell_{1}$-norm of concurrence can be defined as
$$
C^{(1)}(|\psi\rangle)==\sum_{\alpha, \beta=1}^{D_{1},D_{2}}|C_{\alpha\beta}|
=\sum\limits_{i,j=1}^{m}\sqrt{\lambda_{i}\lambda_{j}}.
$$
This expression is nothing but the negativity defined by
\be\label{N}
N\left(|\psi\rangle\right)  =(\left\Vert \rho_{T_{A}}\right\Vert _{1}-1)=(Tr(\rho_{T_{A}}
\rho_{T_{A}}^{\dagger})^{1/2}-1),
\ee
where $T_A$ stands for
the partial transposition with respect to the subsystem $A$, $\left\Vert .\right\Vert _{1}$ is the trace norm.
It is well-known that he negativity is an entanglement monotone.

\section{Bounds for the usual multipartite negativity}

Generalizing the entanglement measure to multipartite quantum states,
we first define the (usual) multipartite entanglement measures, that is, the sum of all the entanglement between
any two subsystems. For simplicity, we only discuss the
tripartite case. But our results can be directly generalized to arbitrary $N$-partite states.

Given a tripartite state $\rho=\ket{\psi}\bra{\psi}$,
$\ket{\psi}\in\mathcal{H}_{A}\otimes\mathcal{H}_{B}\otimes\mathcal{H}_{C}$.
Let $\gamma|\gamma^{^{\prime}}$ denote a bipartition, e.g., $A|BC$.
The usual multipartite concurrence reads
$$
C^{2}(\ket{\psi})=\sum\limits_{\gamma}C^{2}_{\gamma}(\ket{\psi})=\sum\limits_{\gamma=A,B,C}2(1-\tr(\rho^{2}_{\gamma})),
$$
where $\rho_{\gamma}$ are the corresponding reduced density matrices with respect to the subsystem $\gamma$.
$1-\tr(\rho^{2}_{\gamma})=C^{2}_{\gamma}(\ket{\psi})$ is just the linear entropy:
$C^2_{A}(|\psi\rangle)=\sum\limits_{\alpha,\beta}|\langle\psi|J_{\alpha\beta}^{1|23}|\psi^*\rangle|^2$,
$C^2_{B}(|\psi\rangle)=\sum\limits_{\alpha,\beta}|\langle\psi|J_{\alpha\beta}^{2|13}|\psi^*\rangle|^2$,
$C^2_{C}(|\psi\rangle)=\sum\limits_{\alpha,\beta}|\langle\psi|J_{\alpha\beta}^{3|12}|\psi^*\rangle|^2$,
where the operators $J_k$ are defined as for bipartite case before, but correspond to different bipartitions. For instance,
$J^{1|23}_{\alpha,\beta}=L^1_\alpha \otimes S_\beta^{23}$, with $L^1_\alpha $ the $SO(d)$ generators
on $H_{A}$, $S_\beta^{23} $ the $SO(d^2)$ generators on $H_{B}\otimes H_{C}$. $J^{2|13}$ and $J^{3|12}$ are defined in a similar way.

Similarly, we can define the usual multipartite negativity for a multipartite
state $\rho$. For a $d\otimes d \otimes d$ pure state $\rho$,
the usual multipartite negativity reads
\begin{equation}
N(\rho)=\sum\limits_{\gamma}N_{\gamma}(\rho)=2[N_{A}(\rho)+N_{B}(\rho)+N_{C}(\rho)],
\label{multipartitenegativity}
\end{equation}
where $N_{\gamma}(\rho)$ are defined by
\begin{equation}\ba{l}
N_{A}(|\psi\rangle)=\sum\limits_{\alpha,\beta}|\langle\psi|J_{\alpha\beta}^{1|23}|\psi^*\rangle|\\[4mm]
N_{B}(|\psi\rangle)=\sum\limits_{\alpha,\beta}|\langle\psi|J_{\alpha\beta}^{2|13}|\psi^*\rangle|\\[4mm]
N_{C}(|\psi\rangle)=\sum\limits_{\alpha,\beta}|\langle\psi|J_{\alpha\beta}^{3|12}|\psi^*\rangle|.
\label{sorep-negativity}
\ea
\end{equation}

We discuss now the bound for the usual multipartite negativity of superposition.
Let $\mathcal{H}_{A}$, $\mathcal{H}_{B}$ and $\mathcal{H}_{C}$ be the Hilbert spaces of dimension
$d$. We consider two states $|\psi\rangle,|\phi\rangle\in\mathcal{H}_{A}\otimes\mathcal{H}_{B}\otimes\mathcal{H}_{C}$,
$|\psi\rangle=\sum\limits_{1\leq i,j,k\leq d}\psi_{ijk}|ijk\rangle$ and $|\phi
\rangle=\sum\limits_{1\leq i,j,k\leq d}\phi_{ijk}|ijk\rangle$. A superposition of
$|\psi\rangle$ and $|\phi\rangle$ is defined by
$a|\psi\rangle+b|\phi\rangle$, where $|a|^{2}+|b|^{2}=1$.

From Eqs. (\ref{multipartitenegativity}) and (\ref{sorep-negativity}), for
a generic pure state $|\chi\rangle=\sum_{1\leq i,j,k\leq d}\gamma_{ij}%
|ijk\rangle$, we have
\begin{equation}
N(\chi)=\sum\limits_{\gamma}N_{\gamma}(\chi)=\sum\limits_{\gamma}\sum\limits_{\alpha,\beta}|\langle\chi|J_{\alpha\beta}^{\gamma|\bar{\gamma}}|\chi^*\rangle|,
\label{usualne}%
\end{equation}
where $J_{\alpha\beta}^{\gamma|\bar{\gamma}}$ are the tensor product of  generators of the
corresponding bi-partition $\gamma|\bar{\gamma}$, e.g., when $\gamma=1$,
then $\bar{\gamma}=23$, $J^{1|23}_{\alpha,\beta}=L^1_\alpha \otimes S_\beta^{23}$, $L^1_\alpha $
are the $SO(d)$ generators on $H_{A}$, $S_\beta^{23} $ are the $SO(d^2)$
generators on $H_{B}\otimes H_{C}$.

{\bf [Theorem 1]}~Let $|\psi_{1}\rangle$ and $|\psi_{2}\rangle$ be generic  tripartite pure states. Set
$|\chi\rangle=a_{1}|\psi_{1}\rangle+a_{2}|\psi_{2}\rangle$ with $|a_{1}|^{2}%
+|a_{2}|^{2}=1$. Then
\begin{equation}
||\ket{\chi}||^{2}N(\ket{\chi^{\prime}})\leq F_{11}+F_{22}+2F_{12},
\end{equation}
\be
\ba{l}
\Vert\ket{\chi}\Vert^{2}N(\ket{\chi^{\prime}})\geq\max \{F_{11}-F_{22}-2F_{12},\\[3mm]
~~~~~~~~~-F_{11}+F_{22}-2F_{12},-F_{11}-F_{22}+2F_{12}\},
\ea
\ee
where $||\ket{\chi}||^{2}=\langle\chi|\chi\rangle$, $\ket{\chi^{\prime}}=\frac{1}{||\ket{\chi}||}\ket{\chi}$ is the normalized state,
$F_{11}=|a_{1}|^{2}N(\ket{\psi_{1}})$, $F_{22}=|a_{2}|^{2}N(\ket{\psi_{2}})$,
$F_{12}=|a_{1}a_{2}|\sum\limits_{\gamma}\sum\limits_{\alpha,\beta}|\langle\psi_{1}|J_{\alpha\beta}^{\gamma|\bar{\gamma}}|\psi_{2}\rangle^*|$.

{\sf Proof}~From triangular inequality, we have
$$
\ba{l}
\Vert\ket{\chi}\Vert^{2}N(\ket{\chi^{\prime}})=\Vert\ket{\chi}\Vert^{2}\sum\limits_{\gamma}N_{\gamma}(\ket{\chi^{\prime}})\\
=\sum\limits_{\gamma}\sum\limits_{\alpha,\beta}|\langle\chi|J_{\alpha\beta}^{\gamma|\bar{\gamma}}|\chi^*\rangle|\\
=\sum\limits_{\gamma}\sum\limits_{\alpha,\beta}|\langle(a_{1}|\psi_{1}\rangle+a_{2}|\psi_{1}\rangle)|
J_{\alpha\beta}^{\gamma|\bar{\gamma}}|(a_{1}|\psi_{1}\rangle+a_{2}|\psi_{2}\rangle)^*\rangle|\\
\leq F_{11}+F_{22}+2F_{12}.
\ea
$$
For the lower bounds, we have
$$
\ba{l}
\Vert\ket{\chi}\Vert^{2}N(\ket{\chi^{\prime}})=\Vert\ket{\chi}\Vert^{2}\sum\limits_{\gamma}N_{\gamma}(\ket{\chi^{\prime}})\\
=\sum\limits_{\gamma}\sum\limits_{\alpha,\beta}|\langle\chi|J_{\alpha\beta}^{\gamma|\bar{\gamma}}|\chi^*\rangle|\\
=\sum\limits_{\gamma}\sum\limits_{\alpha,\beta}|\langle(a_{1}|\psi_{1}\rangle+a_{2}|\psi_{1}\rangle)
|J_{\alpha\beta}^{\gamma|\bar{\gamma}}|(a_{1}|\psi_{1}\rangle+a_{2}|\psi_{2}\rangle)^*\rangle|\\
\geq F_{11}-F_{22}-2F_{12}.
\ea
$$
The other two lower bounds can be proved similarly.
\quad $\Box$

\section{Bounds for genuine multipartite entanglement}
We now study the genuine multipartite entanglement measures. It is a
challenging problem to qualify the GME. Although having been intensively studied, see e.g. \cite{Huber,Siewert,Guhne10,Guhne11a}, the problem remains far from
being satisfactorily solved.

A proper measure of GME called GME concurrence has been introduced in \cite{Ma11,Chen}, which can distinguish GME from general  entanglement perfectly.
For a tripartite pure state $\ket{\psi}$, the genuine multipartite entanglement measure, GME-concurrence reads \cite{Ma11}:
$$
\begin{array}{l}
C^{2}_{\rm GME}(\ket{\psi})=\min\limits_{\gamma}C_{\gamma}(\ket{\psi})=\min\limits_{\gamma}
\{1-\tr(\rho^{2}_{\gamma})\}\\[3mm]
~~=\min\limits_{A,B,C}\{1-\tr(\rho^{2}_{A}),1-\tr(\rho^{2}_{B}),1-\tr(\rho^{2}_{C})\}.
\end{array}
$$
By definition, any pure state $\rho$ is biseparable if and only if $C_{\rm GME}(\rho)=0$, and $\rho$ is genuine multipartite entangled
if and only if $C_{\rm GME}(\rho)>0$.

For a tripartite pure state $\ket{\psi}$, the genuine multipartite entanglement negativity can be defined by
\begin{equation}
N_{\rm GME}(\psi)=\min\limits_{A,B,C}\{N_{A}(\rho),N_{B}(\rho),N_{C}(\rho)\},
\label{GME-negativity}\end{equation}
where $N_{\gamma}(\rho)$ are defined by (\ref{sorep-negativity}), with $\gamma=A,B,C$.
It is also easy to see that any pure state $\rho$ is biseparable if and only if $N_{\rm GME}(\rho)=0$, and
$\rho$ is genuine multipartite entangled if and only if $N_{\rm GME}(\rho)>0$.

By Eqs. (\ref{GME-negativity}) and (\ref{sorep-negativity}), for
a generic pure state $|\chi\rangle=\sum_{1\leq i,j,k\leq d}\gamma_{ij}%
|ijk\rangle$, we have
$$
N_{\rm GME}(\ket{\chi})=\min\limits_{\gamma}N_{\gamma}(\ket{\chi})=\min\limits_{\gamma}\sum
\limits_{\alpha,\beta}|\langle\chi|J_{\alpha\beta}^{\gamma|\bar{\gamma}}|\chi\rangle^*|.
$$

{\bf [Theorem 2]}~
Let $|\psi_{1}\rangle$ and $|\psi_{2}\rangle$ be generic  tripartite pure states. $|\chi\rangle=a_{1}|\psi_{1}\rangle+a_{2}|\psi_{2}\rangle$ with $|a_{1}|^{2}%
+|a_{2}|^{2}=1$. We have
\be
\ba{l}
\Vert\ket{\chi}\Vert^{2}N_{\rm GME}(\ket{\chi^{\prime}})\leq\min \{g_{11}+f_{22}+2f_{12}, \\[3mm]
~~~~~~~f_{11}+g_{22}+2f_{12},f_{11}+f_{22}+2g_{12}\},
\ea
\ee
\be
\ba{l}
\Vert\ket{\chi}\Vert^{2}N_{\rm GME}(\ket{\chi^{\prime}})\geq\max \{g_{11}-f_{22}-2f_{12}, \\[3mm]
~~~~~~-f_{11}+g_{22}-2f_{12},-f_{11}-f_{22}+2g_{12}\},
\ea
\ee
where $||\ket{\chi}||^{2}=\langle\chi|\chi\rangle$ and $\ket{\chi^{\prime}}=\frac
{1}{||\ket{\chi}||}\ket{\chi}$ is the normalized state,
$f_{ij}=|a_{i}a_{j}|\max\limits_{\gamma}\sum
\limits_{\alpha,\beta}|\langle\psi_{i}|J_{\alpha\beta}^{\gamma|\bar{\gamma}}|\psi_{j}\rangle^*|$,
$g_{ij}=|a_{i}a_{j}|\min\limits_{\gamma}\sum
\limits_{\alpha,\beta}|\langle\psi_{i}|J_{\alpha\beta}^{\gamma|\bar{\gamma}}|\psi_{j}\rangle^*|$.

{\sf Proof}~By triangular inequality, we have
$$
\ba{l}
\Vert\ket{\chi}\Vert^{2}N_{\rm GME}(\ket{\chi^{\prime}})=\Vert\ket{\chi}\Vert^{2}\min\limits_{\gamma}N_{\gamma}(\ket{\chi^{\prime}})\\
=\min\limits_{\gamma}\sum\limits_{\alpha,\beta}|\langle\chi|J_{\alpha\beta}^{\gamma|\bar{\gamma}}|\chi\rangle^*|\\
=\min\limits_{\gamma}\sum\limits_{\alpha,\beta}|\langle(a_{1}|\psi_{1}\rangle+a_{2}|\psi_{2}\rangle)|
J_{\alpha\beta}^{\gamma|\bar{\gamma}}|(a_{1}|\psi_{1}\rangle+a_{2}|\psi_{2}\rangle)\rangle^*|\\
\leq g_{11}+g_{22}+2g_{12}.
\ea
$$
Similarly,
$$
\ba{l}
\Vert\ket{\chi}\Vert^{2}N_{\rm GME}(\ket{\chi^{\prime}})
=\min\limits_{\gamma}\sum\limits_{\alpha,\beta}|\langle\chi|J_{\alpha\beta}^{\gamma|\bar{\gamma}}|\chi\rangle^*|\\
=\min\limits_{\gamma}\sum\limits_{\alpha,\beta}|\langle(a_{1}|\psi_{1}\rangle+a_{2}|\psi_{2}\rangle)|
J_{\alpha\beta}^{\gamma|\bar{\gamma}}|(a_{1}|\psi_{1}\rangle+a_{2}|\psi_{2}\rangle)\rangle^*|\\
\geq g_{11}-g_{22}-2g_{12}.
\ea
$$

Now we need the following simple facts:
if $b_{i},c_{i},d_{i}$, $i=1,2,3$, are positive real numbers, then
\be\label{p1}
\ba{l}
\min\big\{b_{1}+c_{1}+d_{1}, b_{2}+c_{2}+d_{2}, b_{3}+c_{3}+d_{3}\big\}\\[2mm]
\leq \min\big\{b_{1}, b_{2}, b_{3}\big\}+\max\big\{c_{1}, c_{2}, c_{3}\big\}\\[2mm]
~~+\max\big\{d_{1}, d_{2}, d_{3}\big\}
\ea
\ee
and
\be\label{p2}
\ba{l}
\min\big\{b_{1}-c_{1}-d_{1}, b_{2}-c_{2}-d_{2}, b_{3}-c_{3}-d_{3}\big\}\\[2mm]
\geq \min\big\{b_{1}, b_{2}, b_{3}\big\}-\max\big\{c_{1}, c_{2}, c_{3}\big\}\\[2mm]
~~-\max\big\{d_{1}, d_{2}, d_{3}\big\}.
\ea
\ee
The above inequalities can be proved directly.
Without loss of generality, assume $\min\big\{b_{1}+c_{1}+d_{1}, b_{2}+c_{2}+d_{2}, b_{3}+c_{3}+d_{3}\}=b_{1}+c_{1}+d_{1}$. Then,
for $\min\big\{b_{1}, b_{2}, b_{3}\big\}=b_{1}$,
we have $b_{1}+c_{1}+d_{1}\leq b_{1}+\max\big\{c_{1}, c_{2}, c_{3}\big\}+\max\big\{d_{1}, d_{2}, d_{3}\big\}$.
For $\min\big\{b_{1}, b_{2}, b_{3}\big\}\neq b_{1}$, say $\min\big\{b_{1}, b_{2}, b_{3}\big\}= b_{2}$,
then we have $b_{1}+c_{1}+d_{1}\leq b_{2}+c_{2}+d_{2}\leq b_{2}+\max\big\{c_{1}, c_{2}, c_{3}\big\}+\max\big\{d_{1}, d_{2}, d_{3}\big\}$. Hence,
in any cases inequality (\ref{p1}) holds. Inequality (\ref{p2}) can be proved similarly.

Taking the terms $|a_{1}|^{2}\sum\limits_{\alpha,\beta}|\langle\psi_{1}|J_{\alpha\beta}^{1|23}|\psi_{1}\rangle^*|$,
$|a_{1}|^{2}\sum\limits_{\alpha,\beta}|\langle\psi_{1}|J_{\alpha\beta}^{2|13}|\psi_{1}\rangle^*|$,
$|a_{1}|^{2}\sum\limits_{\alpha,\beta}|\langle\psi_{1}|J_{\alpha\beta}^{3|12}|\psi_{1}\rangle^*|$,
$|a_{2}|^{2}\sum\limits_{\alpha,\beta}|\langle|\psi_{2}|J_{\alpha\beta}^{1|23}||\psi_{2}\rangle^*|$,
$|a_{2}|^{2}\sum\limits_{\alpha,\beta}|\langle|\psi_{2}|J_{\alpha\beta}^{2|13}||\psi_{2}\rangle^*|$,
$|a_{2}|^{2}\sum\limits_{\alpha,\beta}|\langle|\psi_{2}|J_{\alpha\beta}^{3|12}||\psi_{2}\rangle^*|$,
$2|a_{1}a_{2}|\sum\limits_{\alpha,\beta}|\langle\psi_{1}|J_{\alpha\beta}^{1|23}|\psi_{2}\rangle^*|$,
$2|a_{1}a_{2}|\sum\limits_{\alpha,\beta}|\langle\psi_{1}|J_{\alpha\beta}^{2|13}|\psi_{2}\rangle^*|$ and
$2|a_{1}a_{2}|\sum\limits_{\alpha,\beta}|\langle\psi_{1}|J_{\alpha\beta}^{3|12}|\psi_{2}\rangle^*|$
as $b_{1}$, $b_{2}$, $b_{3}$, $c_{1}$, $c_{2}$, $c_{3}$, $d_{1}$, $d_{2}$ and $d_{3}$ respectively, we get the following bounds:
$$
\ba{ll}
\Vert\ket{\chi}\Vert^{2}N_{\rm GME}(\ket{\chi^{\prime}})
&\geq |a_{1}|^{2}N_{\rm GME}(\ket{\psi_{1}})\\[3mm]
&-|a_{2}|^{2}\max\limits_{\gamma}\sum\limits_{\alpha,\beta}|\langle\psi_{1}|J_{\alpha\beta}^{\gamma|\bar{\gamma}}\psi_{1}\rangle^*|\\[3mm]
&-2|a_{1}a_{2}|\max\limits_{\gamma}\sum\limits_{\alpha,\beta}|\langle\psi_{1}|J_{\alpha\beta}^{\gamma|\bar{\gamma}}\psi_{2}\rangle^*|
\ea
$$
and
$$
\ba{ll}
\Vert\ket{\chi}\Vert^{2}N_{GME}(\ket{\chi^{\prime}})
&\leq |a_{1}|^{2}N_{\rm GME}(\ket{\psi_{1}})\\[3mm]
&+|a_{2}|^{2}\max\limits_{\gamma}\sum\limits_{\alpha,\beta}|\langle\psi_{1}|J_{\alpha\beta}^{\gamma|\bar{\gamma}}\psi_{1}\rangle^*|\\[3mm]
&+2|a_{1}a_{2}|\max\limits_{\gamma}\sum\limits_{\alpha,\beta}|\langle\psi_{1}|J_{\alpha\beta}^{\gamma|\bar{\gamma}}\psi_{2}\rangle^*|.
\ea
$$
The other two lower bounds and two upper bounds can be proved in the same way.
\quad $\Box$

To show the tightness of our bounds, we consider the following example, the superposition of GHZ-state and W-state,
$\ket{Z(p,\varphi)} = \sqrt{p}\ket{GHZ} +\sqrt{1-p}\ket{W},~~~~0\leq p\leq 1$, Our upper bounds of the usual multipartite negativity and the GME-negativity
for state $\ket{Z(p,\varphi)}$ are given by $32(1-p)+16\sqrt{6p(1-p)}+24p$ and  $\frac{16}{3}(1-p)+\frac{8}{3}\sqrt{6p(1-p)}+4p$ respectively,
which are just the exact values of the usual multipartite negativity and the GME-negativity.

\section{Conclusion and remarks}
By deriving analytical tight lower and upper
bounds of the usual multipartite negativity and the genuine multipartite entanglement negativity,
we have investigated how the usual and the genuine multipartite entanglement are distributed among the components of superposed quantum states.
The example also shows that our results can be used to study of GME quantification itself.
Above all, our results can be directly generalized to arbitrary $N$-partite quantum states.

\bigskip

{\noindent}\emph{Acknowledgment} Z. Ma is supported by NSF of China (11275131)
and by the Foundation of China Scholarship Council (2010831012). Z. Chen is supported by NSF of China (11201427)
and by the Foundation of China Scholarship Council (201207285006).
S.M. Fei is supported by the NSFC under number 11275131.
Z.Ma Thanks for discussion with S. Severini.

\end{document}